\documentclass[showpacs,amssymb,aps]{revtex4}
\usepackage{graphics}

\begin{document}

\title{General structure of the photon self-energy in non-commutative QED}

\author{F. T. Brandt$^a$, Ashok Das$^b$ and
J. Frenkel$^a$}
\affiliation{$^a$Instituto de F\'{\i}sica,
Universidade de S\~ao Paulo,
S\~ao Paulo, SP 05315-970, BRAZIL}
\affiliation{$^b$Department of Physics and Astronomy,
University of Rochester,
Rochester, NY 14627-0171, USA}


\begin{abstract}
We study the behavior of the photon two point function, in
non-commutative QED, in a general covariant gauge and in arbitrary
space-time dimensions. We show, to all orders, that the photon
self-energy is transverse. Using an appropriate extension of the
dimensional regularization method, we evaluate the one-loop
corrections, which show that the theory is renormalizable. We also
prove, to all orders, that the poles of the photon propagator are gauge
independent and briefly discuss some other related aspects.
\end{abstract}


\maketitle

\section{Introduction}

Non-commutative theories have generated a lot of interest in recent
years \cite{Gonzalez-Arroyo:1983ub,Filk:1996dm,Martin:1999aq,Sheikh-Jabbari:1999iw,Krajewski:1999ja,Bigatti:1999iz,Maldacena:1999mh,Iso:2000ew,Arcioni:1999hw,Minwalla:1999px,Aref'eva:1999sn,Gross:2000ba,Matusis:2000jf,Hayakawa:1999zf,girotti:2000gc,Zanon:2000nq,Das:2001kf,Khoze:2000sy,Liu:2000ad,VanRaamsdonk:2001jd,Armoni:2000xr,Ruiz:2000hu,Bichl:2001cq,foot}.
These are theories defined on a manifold where the coordinates
do not commute. As a result, in such theories, there is a natural
second rank anti-symmetric tensor with the canonical dimension of inverse
mass squared. Several interesting features develop in such
theories. First of all, because of non-commutativity of coordinates,
the natural product of functions, on such a space, is the star product
of Gr\"{o}newold and Moyal \cite{Groenewold:1946,Moyal:1949sk}.
One of the consequences of such a  product
is that, in a physical theory, the interaction terms develop a
momentum dependent phase factor (in momentum space). However, the two
point functions and, therefore, the propagators do not
modify. Furthermore, explicit loop calculations show that the
conventional ultraviolet terms in a commutative theory get distributed
into two parts -- one that is ultraviolet divergent while the other
is ultraviolet finite. However, these ultraviolet finite terms become
singular in the infrared limit and the amplitudes become non-analytic.

All these features are quite fascinating and puzzling, since they are
somewhat similar to what happens in thermal field 
theories \cite{kapusta:book89,lebellac:book96,das:book97}. 
For example, we
know that, at finite temperature, we have a new scale, the
temperature, and an additional, natural Lorentz vector, $u^{\mu}$,
which is the velocity of the heat bath. In a thermal field theory,
however, the interaction vertices do not modify, rather the
propagators do, because of  (anti) periodic boundary conditions. In
this sense, thermal field theories and non-commutative field theories
seem complementary and a natural question of interest is whether there
is any redefinition of variables that may map one to the other. It is
also known that no new ultraviolet divergences develop at finite
temperature. However, the infrared divergences do become more severe,
once again similar to what happens in non-commutative theories,
at least in one loop. Even more
fascinating is the observation that, at finite temperature, amplitudes
become non-analytic at the origin in the energy-momentum plane, which is
reminiscent of the non-analyticity in non-commutative theories. In
thermal field theories, the physical origin of the non-analyticity is
well understood. Namely, in a thermal medium, new channels of reaction
develop leading to new branch cuts, which is the reason for the 
non-analyticity. In the same spirit, it will be interesting to
understand if  there is a physical origin of the non-analyticity in
non-commutative theories.

A lot is already known about thermal field theories and even though
we do not yet know whether non-commutative theories and thermal theories
can be mapped into each other, we may make use of some of the
techniques that have been developed in connection with thermal field
theories, to learn more about non-commutative field theories. It is
with this goal that we take up a systematic study of the photon
self-energy in non-commutative QED in a general covariant gauge in 
arbitrary dimensions. Since the contribution of the fermion loop to the
photon self-energy has been studied in detail in the past, we 
concentrate only on the contributions coming from internal gauge and
ghost loops. Our study leads to a number of interesting
features that bring out similarities and differences between thermal
field theories and non-commutative field theories. For example, we
find that  although non-commutative QED has a non-Abelian character
because of the star product, the self-energy is transverse to all
orders in a general covariant gauge in any dimension. This has to be
contrasted with the self-energy of thermal QCD, which is, in general,
not  transverse. We verify this all orders result
by explicitly calculating the self-energy at one loop. The
calculations in a gauge theory are, of course, best carried out in
dimensional regularization. However, most of the calculations in
non-commutative theories, so far, have used the method due to
Schwinger (the difficulty is mainly due to the exponential phase
factor). Therefore, as a first step, we have generalized the formulae
of dimensional regularization to non-commutative theories and this
indeed simplifies the calculations quite a bit. The explicit
calculation, at one loop, shows that the self-energy is gauge
dependent, but does not develop any new kind of ultraviolet
divergence, so that the theory is renormalizable. Furthermore, in the
infrared limit, the
imaginary part of the contributions to the self-energy, coming from
these graphs, identically vanishes.  This is, in
fact, interesting in that the non-analyticity in the non-commutative
QED does not seem to be connected with new imaginary parts in the
amplitude. Away from the infrared, however, the self-energy does have
imaginary parts which are necessary for unitarity. 
The gauge dependence of the
self-energy raises the question of the behavior of the poles of the photon
propagator in this theory and we prove to all orders,
using the Nielsen identity \cite{das:book97,Nielsen:1975fs}, 
that, in spite of the gauge dependence of
the self-energy, the poles of the propagator are gauge independent.

The paper is organized as follows. In section {\bf II}, we briefly
review the structure of non-commutative QED. In section {\bf III},
drawing from previous experience with thermal field theories, we show that
the photon self-energy, in this theory, is transverse to all orders in
a general covariant gauge in any dimension. In section {\bf IV}, we
generalize the formulae of dimensional regularization to
non-commutative theories. One loop calculations, using dimensional
regularization, are presented in section {\bf V}, where we also talk
about various aspects of the result. In section {\bf VI}, we analyze
the imaginary part of the photon self-energy and bring out some
interesting features associated with it. In section {\bf VII}, we
prove, using the Nielsen identity, that the poles of the
photon propagator are gauge independent to all orders. We present a
short conclusion in section {\bf VIII} and give details on the
derivation of the Nielsen identity in the appendix.

\section{Non-commutative QED}

Non-commutative QED differs from the conventional commutative QED in
the following
manner. First of all, the theory is defined on a manifold, where the
coordinates do not commute. Rather, they satisfy
\begin{equation}
[x^{\mu}, x^{\nu}] = i \theta^{\mu\nu}
\end{equation}
where $\theta^{\mu\nu} = -\theta^{\nu\mu}$ has the canonical
dimension of inverse mass squared. To avoid problems with unitarity,
we will assume that only the space-space components of
$\theta^{\mu\nu}$ are nonzero, namely, that the time coordinate 
commutes with all the coordinates. An immediate consequence of the
non-commutativity of coordinates is that products of functions on this
manifold are naturally defined by the Gr\"{o}newold-Moyal star product
\begin{equation}
f(x)\star g(x) = \left[e^{{i\over
2}\theta^{\mu\nu}\partial_{\mu}^{(\zeta)}\partial_{\nu}^{(\eta)}}
f(x+\zeta) g(x+\eta)\right]_{\zeta=0=\eta}
\end{equation}
The star product also naturally introduces a Moyal bracket of two
bosonic functions as
\begin{equation}
[f,g]_{\rm MB} = f\star g - g\star f
\end{equation}

With these, we can define the action for non-commutative QED as
\begin{equation}
S_{\rm inv} = \int d^{n}x\,{\cal L}_{\rm inv} = \int
d^{n}x\,\left(-{1\over 4} F_{\mu\nu}\star F^{\mu\nu}  +
\bar{\psi}\star (iD\!\!\!\!\slash - m)\psi\right)
\end{equation}
where $n$ is the number of space-time dimensions and
\begin{eqnarray}
D_{\mu}\psi & = & \partial_{\mu}\psi - i e A_{\mu}\star
\psi\nonumber\\
F_{\mu\nu} & = & \partial_{\mu}A_{\nu} - \partial_{\nu}A_{\mu} -
ie[A_{\mu},A_{\nu}]_{\rm MB}
\end{eqnarray}
This action can be easily verified to be invariant under the gauge
transformations
\begin{eqnarray}
\psi (x) & \rightarrow & \psi' (x) = U(x)\star \psi (x)\nonumber\\
A_{\mu} (x) & \rightarrow & A_{\mu}' (x) = U(x)\star A_{\mu} (x) \star
U^{-1} (x) + {i\over e} U(x)\star \partial_{\mu}U^{-1}(x)
\end{eqnarray}
Infinitesimally, the transformations take the form
\begin{eqnarray}
\delta \psi(x) & = & i\epsilon(x)\star \psi(x)\nonumber\\
\delta A_{\mu} & = & {1\over e}\,D_{\mu}\epsilon(x) = {1\over
e}\left(\partial_{\mu}\epsilon - ie \left[A_{\mu},\epsilon\right]_{\rm
MB}\right)
\end{eqnarray}
where $\epsilon(x)$ is the parameter of infinitesimal
transformations. We can, of course, add to this action a gauge fixing
and a ghost action. For covariant gauge fixing, they will have the form
\begin{equation}
S_{\rm gf} + S_{\rm ghost} = \int d^{n}x\,\left(-{1\over 2\xi}
(\partial_{\mu}A^{\mu})\star (\partial_{\nu}A^{\nu}) +
\partial^{\mu}\overline{c}\star (\partial_{\mu}c - ie [A_{\mu},c]_{\rm
MB})\right)
\end{equation}
where $\xi$ is the gauge fixing parameter. We note, therefore, that we
can  write the complete action for non-commutative QED in a general
covariant gauge as
\begin{equation}
S = S_{\rm inv} + S_{\rm gf} + S_{\rm ghost}\label{action}
\end{equation} 

Thus, we see that because of the star product, the action in
(\ref{action}),  for non-commutative QED, has a non-Abelian structure
through the Moyal bracket. The star product has 
some interesting consequences. In particular, under an integral, the star
product of two functions is the same as an ordinary product (namely, the
difference between the two integrands is a total divergence that
integrates to zero for functions
with appropriate asymptotic fall off). Similarly, the star product of
any number of functions, under an integral, satisfies cyclicity. As a
result of these, it follows that the two point
functions and the propagators of a non-commutative field  theory are
the same  as their commutative counterparts. However, the
interaction vertices have an exponential dependence on
$\theta^{\mu\nu}$ as well as the momenta carried by the fields. Thus,
for example, for the action in (\ref{action}), the Feynman rules for the theory
are as follows. First, the propagators of the theory are 
\begin{eqnarray}
\includegraphics*{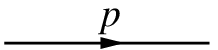} & : \;\;\; & \;\;\;
\displaystyle{{i\over p\!\!\!\slash - m + i\epsilon} = iS(p)}\nonumber\\
\includegraphics*{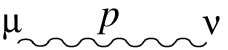}
& : \;\;\;&\;\;\; -\displaystyle{{i\over (p^{2}+i\epsilon)} \left(\eta_{\mu\nu} -
 (1-\xi){p_{\mu}p_{\nu}\over p^{2}}\right) = iD_{\mu\nu}(p)}\nonumber\\
\includegraphics*{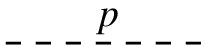}
 & : \;\;\;&\;\;\; 
\displaystyle{{i\over p^{2} + i\epsilon} = iD(p)}\label{propagator}
\end{eqnarray}
These are the same as in the commutative theory. Introducing the
notation, 
\begin{equation}
p\times q = \theta^{\mu\nu}p_{\mu}q_{\nu}
\end{equation}
the vertices, on the other hand, have the following forms (with the
momentum conserving delta functions omitted)
\begin{eqnarray}
\begin{array}{c}
\includegraphics*{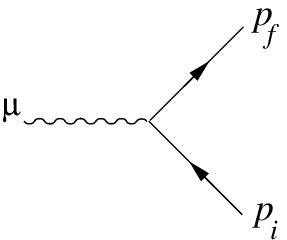}
\end{array}
 & :\;\;\;  & \;\;\; 
ie\gamma^{\mu} e^{{i\over 2} p_{i}\times p_{f}}\nonumber\\  & & \nonumber\\
\begin{array}{c}\includegraphics*{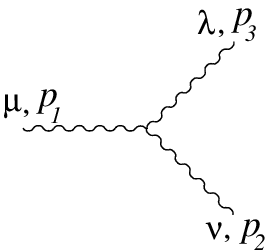}\end{array}
 & :\;\;\;  &\;\;\;  -2i\, e \sin {p_{1}\times p_{2}\over
 2}\left[(p_{1}-p_{2})^{\lambda}\eta^{\mu\nu} +
 (p_{2}-p_{3})^{\mu}\eta^{\nu\lambda} +
 (p_{3}-p_{1})^{\nu}\eta^{\lambda\mu}\right]\nonumber\\  & & \nonumber\\
\begin{array}{c}\includegraphics*{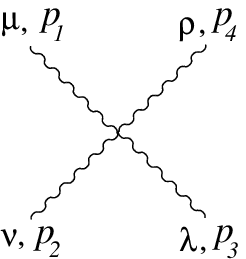}\end{array}
 & :\;\;\;  &\;\;\;  - 4\, e^{2}\left[(\eta^{\mu\lambda}\eta^{\nu\rho} -
 \eta^{\mu\rho}\eta^{\nu\lambda}) \sin {p_{1}\times p_{2}\over 2} \sin
 {p_{3}\times p_{4}\over 2}\right.\nonumber\\
 &  & \qquad\quad + (\eta^{\mu\rho}\eta^{\nu\lambda} -
 \eta^{\mu\nu}\eta^{\lambda\rho}) \sin {p_{3}\times p_{1}\over 2} \sin
 {p_{2}\times p_{4}\over 2}\nonumber\\ & & \nonumber\\ & & \nonumber\\
 &  & \qquad\quad\left. + (\eta^{\mu\nu}\eta^{\lambda\rho} -
 \eta^{\mu\lambda}\eta^{\nu\rho}) \sin {p_{1}\times p_{4}\over 2} \sin
 {p_{2}\times p_{3}\over 2}\right]\nonumber\\& & \nonumber\\ & & \nonumber\\
\begin{array}{c}\includegraphics*{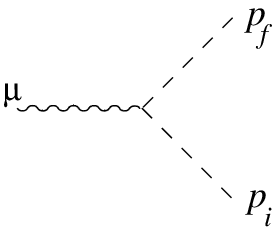}\end{array}
 & :\;\;\;  &\;\;\;  2ie p_{f}^{\mu} \sin {p_{i}\times p_{f}\over 2}\label{vertex}
\end{eqnarray}

In this paper, we will study, systematically, the photon self-energy,
at one loop, in a general covariant gauge in an arbitrary dimension. Since
the fermion contribution to this is the same as in a commutative
theory (namely, the diagram is `` planar''), and has been studied in
the literature, we will concentrate
only on the other graphs, which do not occur in  commutative
QED. The star product, of course, introduces a non-Abelian structure in this
theory. But more than that, in such theories, we have  an independent Lorentz
structure which can, in principle, introduce a 
behavior parallel to that at finite temperature. For example, in the
self-energy, there 
is one independent external momentum so that we can think of
$\bar{u}^{\mu} = \theta^{\mu\nu}p_{\nu}$ as being analogous to the component
of the velocity of the heat bath perpendicular to momentum at finite
temperature. A lot is known
about  the self-energy of commutative
Yang-Mills theory at finite temperature and  our goal is to
exploit the known features of such studies to understand the behavior
of the self-energy in non-commutative QED in a general covariant
gauge in any dimension. Even though our actual calculations are at
one loop,  in the process, we will find some interesting all orders
results for the self-energy as well as the propagator in non-commutative QED.

\section{Transversality of the polarization tensor}

It is well known in commutative QCD that, at finite temperature, the
self-energy  is not transverse in a general covariant 
gauge \cite{kobes:1989up,brandt:1997se} (It is
transverse at one loop, only in the Feynman gauge). In spite of the 
apparent similarity of non-commutative theories with thermal field
theories,  we will show in the
following that the photon self-energy in non-commutative QED is transverse to
all orders in any covariant gauge and in any dimension, which is the
behavior of the gauge self-energy in commutative QCD at zero temperature.

To show this, let us introduce some techniques from finite temperature
field theory \cite{Weldon:1996kb}.
Consider a theory with a natural vector $u^{\mu}$ (for
example, the velocity of the heat bath at finite temperature). Then,
given a momentum vector $p^{\mu}$, let us define the component of
$u^{\mu}$ orthogonal to $p^{\mu}$ as
\begin{equation}
\bar{u}^{\mu} = \left(u^{\mu} - {(u\cdot p)\over
p^{2}}\,p^{\mu}\right),\qquad p_{\mu}\bar{u}^{\mu} = 0
\end{equation}
In such a theory, the self-energy for the gauge field will have the
most general, all orders decomposition given by
\begin{equation}
\Pi^{\mu\nu} = A\, \left(\eta^{\mu\nu} - {p^{\mu}p^{\nu}\over
p^{2}}\right) + B\, {\bar{u}^{\mu}\bar{u}^{\nu}\over \bar{u}^{2}} +
C\, {p^{\mu}\bar{u}^{\nu} + p^{\nu}\bar{u}^{\mu}\over \bar{u}^{2}} +
D\,{p^{\mu}p^{\nu}\over p^{2}}\label{decomp}
\end{equation}
It follows from this that
\begin{equation}
p_{\mu}\Pi^{\mu\nu} = {Cp^{2}\over \bar{u}^{2}}\,\bar{u}^{\nu} +
Dp^{\nu}\neq 0
\end{equation}
unless $C = 0 = D$.

Adding the tree level two point function, we can write the complete two point
function to all orders as
\begin{eqnarray}
\Gamma^{\mu\nu} & = & \widehat{\Gamma}^{\mu\nu} + {p^{\mu}p^{\nu}\over
\xi}\nonumber\\
 & = & \left[(p^{2} + A)\, \left(\eta^{\mu\nu} - {p^{\mu}p^{\nu}\over
p^{2}}\right) + B\,{\bar{u}^{\mu}\bar{u}^{\nu}\over \bar{u}^{2}} +
C\,{p^{\mu}\bar{u}^{\nu}+p^{\nu}\bar{u}^{\mu}\over \bar{u}^{2}} +
D\,{p^{\mu}p^{\nu}\over p^{2}}\right] + {p^{\mu}p^{\nu}\over \xi}
\end{eqnarray}
The complete propagator is defined to be
\begin{equation}
\Gamma^{\mu\nu} D_{\nu\lambda} = -\delta^{\mu}_{\lambda}
\end{equation}
and satisfies, in consequence of the Slavnov-Taylor 
identity \cite{Taylor:ff},
\begin{equation}
p^{\mu}p^{\nu} D_{\mu\nu} = -\xi
\end{equation}
Let us now define
\begin{equation}
G_{\mu} = p^{\nu}D_{\mu\nu},\qquad p^{\mu}G_{\mu} = -\xi
\end{equation}
Then, it follows, in a simple manner, that
\begin{equation}
\widehat{\Gamma}^{\mu\nu} G_{\nu} = \left(\Gamma^{\mu\nu} -
{p^{\mu}p^{\nu}\over \xi}\right) p^{\rho}D_{\nu\rho} = 
-p^{\mu} + p^{\mu} = 0
\end{equation}
This shows that $\widehat{\Gamma}^{\mu\nu}$ has a zero mode, which can
be explicitly constructed to be
\begin{equation}
G^{0}_{\mu} =  -{\xi\over p^2}
\left(p_{\mu} - {D\over C}\,\bar{u}_{\mu}\right)\label{zeromode}
\end{equation}
with
\begin{equation}
D = {p^{2} C^{2}\over \bar{u}^{2} (p^{2} +A + B)}\label{zero}
\end{equation}

Thus far, our analysis has been quite general. Let us next turn to
non-commutative QED. In this case, we can identify
\begin{equation}
\bar{u}^{\mu} = \theta^{\mu\nu}p_{\nu}
\end{equation}
Furthermore, from the Feynman rules, let us note that the self-energy
diagrams (involving internal photon and ghost lines) are invariant
under
\[
\theta^{\mu\nu} \rightarrow - \theta^{\mu\nu}
\]
Therefore, the self-energy must be an even function of
$\bar{u}^{\mu}$. (We note here that this property holds even with
fermion interactions and is a consequence of charge conjugation
invariance of the two point function \cite{Sheikh-Jabbari:2000vi}.)
It follows, then, that the
coefficient, $C$, in (\ref{decomp})
must be odd in $\bar{u}^{\mu}$. Since it is a scalar function, the most
general form it can have is
\begin{equation}
C = (\bar{u}\cdot p)\,E = 0
\end{equation}
where $E$ denotes a scalar function, even in $\bar{u}^{\mu}$, and we
have used the fact that $\bar{u}^{\mu}$ is orthogonal to
$p^{\mu}$. Since $C=0$, it follows from (\ref{zero}) that $D=0$. 

Therefore, to all orders, we determine that the most general form for the
self-energy  in non-commutative
QED, in a general covariant gauge in any dimension, can be written as
\begin{equation}
\Pi^{\mu\nu} = A\,\left(\eta^{\mu\nu} - {p^{\mu}p^{\nu}\over
p^{2}}\right) + B\,{\bar{u}^{\mu}\bar{u}^{\nu}\over
\bar{u}^{2}}\label{selfenergy}
\end{equation}
which is manifestly transverse. (Note that, in this case, the zero
mode in (\ref{zeromode}) simply reduces to $-\xi p_\mu/p^2$.) 
The coefficients
$A$ and $B$, of course, will be dependent on the gauge fixing
parameter, unlike in commutative QED, and we would like to evaluate
these functions at one loop.

\section{Dimensional regularization in non-commutative field theory}

The calculations in non-commutative field theory have so far been
mostly carried out using the methods of Schwinger. However, from
studies in 
commutative gauge field theories, we know that dimensional
regularization \cite{'thooft:1972fi}
is extremely simple and powerful which, while
maintaining gauge invariance, allows the proof of many results in a
natural manner. Therefore, it is quite useful to try to extend the
method of dimensional regularization to non-commutative 
theories \cite{Armoni:2000xr,Ruiz:2000hu}.
In what follows, we will derive the dimensional regularization formulae,
relevant to non-commutative theories in two different ways, both
leading to the same results.

In non-commutative theories (for example, in non-commutative QED, see
(\ref{vertex})), the interactions involve a momentum
dependent phase factor. Therefore, a generic loop integral, that arises in
such theories, has the form
\begin{equation}
I = \int {d^{n}k\over (2\pi)^{n}}\, {e^{i\bar{\theta}\cdot k}\over
(k^{2} - M^{2})^{\alpha}}
\end{equation}
where we have defined
\begin{equation}
\bar{\theta}^{\mu} = \theta^{\mu\nu}f_{\mu}(p_{1},\cdots , p_{n})
\end{equation}
with $f_{\mu}$ representing a function of the external momenta,
$p_{1},\cdots ,p_{n}$,  and
$M^{2}$ represents the term that arises from combining denominators
using the Feynman parameters and shifting. $M^{2}$, in general, depends on the
external  momenta, masses of the theory as well as
the Feynman parameters. For simplicity, we have ignored writing the
integration over the Feynman parameters that needs to be carried out. To
evaluate this integral, we first rotate to Euclidean space, so that we
have (note that with our choice of $\theta^{\mu\nu}$ having only space
indices, the exponent does not change sign upon rotation to Euclidean
space) 
\begin{equation}
I = i (-1)^{\alpha} \int {d^{n}k\over (2\pi)^{n}}\,
{e^{i\bar{\theta}\cdot k}\over (k^{2}+M^{2})^{\alpha}}
\end{equation}

It is the momentum dependent exponential that seems
formidable. However, it turns out that 
it is not hard to evaluate integrals of this kind and let us
present two different, but equivalent ways that lead to the same
result. First of all, let us decompose the vector $k_{\mu}$ to
longitudinal and transverse components with respect to
$\bar{\theta}_{\mu}$. Let us introduce the decomposition,
\begin{equation}
k_{\|\,\mu} = {\bar{\theta}\cdot k\over
\bar{\theta}^{2}}\,\bar{\theta}_{\mu},\qquad k_{\bot\,\mu} =
k_{\mu}-k_{\|\,\mu}
\end{equation}
so that
\begin{equation}
\bar{\theta}\cdot k_{\|} = \bar{\theta}\cdot k,\qquad
\bar{\theta}\cdot k_{\bot} = 0
\end{equation}

In terms of these components, then, we can write
\begin{eqnarray}
I & = & {i (-1)^{\alpha}\over (2\pi)^{n}} \int
dk_{\|}\,e^{i\bar{\theta}\cdot k_{\|}} \int d^{n-1}k_{\bot}
\,{1\over (k_{\bot}^{2}+k_{\|}^{2} +
M^{2})^{\alpha}}\nonumber\\
 & = & {i (-1)^{\alpha}\over (2\pi)^{n}} \int_{-\infty}^{\infty}
dk_{\|}\,{\pi^{n-1\over 2}\Gamma(\alpha - {n-1\over 2})\over
\Gamma(\alpha)} {e^{i\bar{\theta}\cdot k_{\|}}\over (k_{\|}^{2} +
M^{2})^{\alpha - {n-1\over 2}}}\nonumber\\
 & = & {2i(-1)^{\alpha}\over (2\pi)^{n}}\, {\pi^{n-1\over 2}\Gamma
(\alpha - {n-1\over 2})\over \Gamma (\alpha)} \int_{0}^{\infty}
dk_{\|}\, {\cos \bar{\theta}\cdot k_{\|}\over (k_{\|}^{2} +
M^{2})^{\alpha -{n-1\over 2}}} 
\end{eqnarray}

Recalling that $k_{\|\,\mu}$ is parallel to $\bar{\theta}_{\mu}$, we
note that $\cos \bar{\theta}\cdot k_{\|} = \cos |\bar{\theta}|k_{\|}$
where $|\bar{\theta}| = (-\bar{\theta}_{\mu}\bar{\theta}_{\mu})^{1\over
2}$ and $k_{\|} = (k_{\|\,\mu}k_{\|\,\mu})^{1\over 2}$. Therefore,
defining a new variable $x= |\bar{\theta}|k_{\|}$, we can write the
integral as
\begin{eqnarray}
I  & = & {2i(-1)^{\alpha}\over (2\pi)^{n}}\,{\pi^{n-1\over 2}\Gamma(\alpha
-{n-1\over 2})\over \Gamma(\alpha)}\,|\bar{\theta}|^{2\alpha
-n}\int_{0}^{\infty} dx\,{\cos x\over (x^{2} +
(|\bar{\theta}|M)^{2})^{\alpha - {n-1\over 2}}}\nonumber\\
 & = & 2i(-1)^{\alpha} {\pi^{n-1\over 2}\over
(2\pi)^{n}}\,{\Gamma(\alpha -{n-1\over 2})\over
\Gamma(\alpha)}\,|\bar{\theta}|^{2\alpha -n}\times
{(2|\bar{\theta}|M)^{{n\over 2}-\alpha}\over \sqrt{\pi}}\,\cos
\pi({n\over 2}-\alpha) \Gamma({n+1\over 2}-\alpha) K_{\alpha -{n\over
2}} (|\bar{\theta}|M)\nonumber\\
 & = & i (-1)^{\alpha} {\pi^{n\over 2}\over (2\pi)^{n}}\,{1\over
\Gamma(\alpha)}\,{1\over (M^{2})^{\alpha -{n\over 2}}}\,2
\left({|\bar{\theta}|M\over 2}\right)^{\alpha -{n\over 2}}\,K_{\alpha -
{n\over 2}} (|\bar{\theta}|M)\label{dimreg}
\end{eqnarray}
Here $K_{\alpha}$ denotes the Bessel function and, in the intermediate
steps, we have used some identities involving the gamma 
functions  \cite{gradshteyn}.
It is worth pointing out here that the identification of the last
integral with the Bessel function is strictly valid when $n<2\alpha
+1$, where it is straightforward to show, with the help of standard
tables that
\begin{equation}
\lim_{z\rightarrow 0}\, z^{\nu} K_{\nu} (z) \longrightarrow 2^{\nu -1}
\Gamma (\nu)\label{convgt}
\end{equation}
so that we have
\begin{equation}
\lim_{\bar{\theta}\rightarrow 0}\, I \rightarrow i (-1)^{\alpha}
{\pi^{n\over 2}\over (2\pi)^{n}}\,{\Gamma (\alpha -{n\over 2})\over
\Gamma(\alpha)}\,{1\over (M^{2})^{\alpha -{n\over 2}}}
\end{equation}
However, as in dimensional regularization in commutative
theories, we analytically continue this result to other dimensions.

Let us next give an alternate derivation of the dimensional
regularization formula for our basic integral. We note that
\begin{eqnarray}
I & = & i(-1)^{\alpha} \int {d^{n}k\over
(2\pi)^{n}}\,{e^{i\bar{\theta}\cdot k}\over
(k^{2}+M^{2})^{\alpha}}\nonumber\\
 & = & {i(-1)^{\alpha}\over \Gamma(\alpha)}\,\left(-{\partial\over
\partial M^{2}}\right)^{\alpha -1}\int {d^{n}k\over
(2\pi)^{n}}\,{e^{i\bar{\theta}\cdot k}\over k^{2}+M^{2}}\nonumber\\
 & = & {i(-1)^{\alpha}\over \Gamma(\alpha)}\left(-{\partial\over
\partial M^{2}}\right)^{\alpha -1} \int {d^{n}k\over
(2\pi)^{n}}\int_{0}^{\infty} d\lambda\,e^{-\lambda (k^{2}+M^{2}) +
i\bar{\theta}\cdot k}\nonumber\\
 & = & {i(-1)^{\alpha}\pi^{n\over 2}\over (2\pi)^{n}}{1\over
\Gamma(\alpha)}\int_{0}^{\infty} d\lambda\,\lambda^{\alpha -1-{n\over
2}}\,e^{-\lambda M^{2} - {|\bar{\theta}|^{2}\over
4\lambda}}\nonumber\\
 & = & i (-1)^{\alpha} {\pi^{n\over 2}\over (2\pi)^{n}}\,{1\over
\Gamma(\alpha)}\,{1\over (M^{2})^{\alpha -{n\over
2}}}\,2\left({|\bar{\theta}|M\over 2}\right)^{\alpha -{n\over
2}}\,K_{\alpha -{n\over 2}} (|\bar{\theta}|M)
\end{eqnarray}
which is exactly the same formula as derived earlier in (\ref{dimreg}).

This, therefore, generalizes dimensional regularization to
non-commutative theories and evaluates the basic integral that arises
in a non-commutative field theory. In the study of self-energy in
gauge theories,
however, we need integrals involving additional tensor structures,
which can  be evaluated in
the following simple manner. First, let us note that, if we are
interested in the self-energy, there is only one independent external
momentum and, therefore, 
\begin{equation}
 \bar{\theta}_{\mu} = \theta_{\mu\nu}p_{\nu}
\end{equation} 
where $p_{\mu}$ is the external momentum. Second, in this case,
$M^{2}=  x(1-x)p^{2}$, where
$x$ is the Feynman parameter that arises in combining two
denominators. Now, if we introduce an auxiliary vector $z_{\mu}$,
then, we obtain, following our earlier derivation in (\ref{dimreg}), that
\begin{equation}
I(z) = i(-1)^{\alpha} \int {d^{n}k\over
(2\pi)^{n}}\,{e^{i(\bar{\theta}+z)\cdot k}\over
(k^{2}+M^{2})^{\alpha}} = i(-1)^{\alpha} {\pi^{n\over 2}\over
(2\pi)^{n}}\,{1\over\Gamma(\alpha) (M^{2})^{\alpha -{n\over
2}}}\,2\left({|\bar{\theta} + z|M\over 2}\right)^{\alpha -{n\over
2}}\,K_{\alpha -{n\over 2}} (|\bar{\theta}+z|M)
\end{equation}
For small values of $z^{\mu}$, expanding in a Taylor series and using
identities involving the Bessel
functions, this result would generate the integrals
involving all other tensor structures.

For completeness, we list below the Euclidean forms of the integrals
that we will need in the subsequent sections.
\begin{eqnarray}
\int {d^{n}k\over (2\pi)^{n}}\,{e^{i\bar{\theta}\cdot k}\over
(k^{2}+M^{2})^{\alpha}} & = & {\pi^{n\over 2}\over
(2\pi)^{n}}\,{1\over \Gamma(\alpha)}\,{1\over (M^{2})^{\alpha -
{n\over 2}}}\,2\left({|\bar{\theta}|M\over 2}\right)^{\alpha -{n\over
2}}\, K_{\alpha -{n\over 2}} (|\bar{\theta}|M)\nonumber\\
\int {d^{n}k\over (2\pi)^{n}}\,{e^{i\bar{\theta}\cdot k}\,
(\bar{\theta}\cdot k)^{2}\over (k^{2} + M^{2})^{\alpha}} & = &
{\pi^{n\over 2}\over (2\pi)^{n}}\,{1\over
\Gamma(\alpha)}\,{|\bar{\theta}|^{2}\over (M^{2})^{\alpha -1-{n\over
2}}}\left[(2\alpha-1-n)\left({|\bar{\theta}|M\over
2}\right)^{\alpha-1-{n\over 2}} K_{\alpha-1-{n\over 2}}
(|\bar{\theta}|M)\right.\nonumber\\
 &  & \qquad\qquad\qquad\qquad\left. - 2\left({|\bar{\theta}|M\over
2}\right)^{\alpha-{n\over 2}} K_{\alpha-{n\over 2}}
(|\bar{\theta}|M)\right]\nonumber\\
\int {d^{n}k\over (2\pi)^{n}}\,{e^{i\bar{\theta}\cdot k}\,
k_{\mu}k_{\nu}\over (k^{2}+M^{2})^{\alpha}} & = & A_{\alpha}
\delta_{\mu\nu} + B_{\alpha}
{\bar{\theta}_{\mu}\bar{\theta}_{\nu}\over \bar{\theta}^{2}}\label{dimreg1}
\end{eqnarray}
where
\begin{eqnarray}
A_{\alpha} & = & {\pi^{n\over 2}\over (2\pi)^{n}}\,{1\over
\Gamma(\alpha)}\, {1\over (M^{2})^{\alpha-1-{n\over
2}}}\,\left({|\bar{\theta}|M\over 2}\right)^{\alpha-1-{n\over 2}}
K_{\alpha-1-{n\over 2}} (|\bar{\theta}|M)\nonumber\\
B_{\alpha} & = & {\pi^{n\over 2}\over (2\pi)^{n}}\,{1\over
\Gamma(\alpha)}\, {1\over (M^{2})^{\alpha-1-{n\over
2}}}\,\left[(2\alpha-2-n)\left({|\bar{\theta}|M\over
2}\right)^{\alpha-1-{n\over 2}} K_{\alpha-1-{n\over 2}}
(|\bar{\theta}|M) - 2\left({|\bar{\theta}|M\over 2}\right)^{\alpha
-{n\over 2}} K_{\alpha -{n\over 2}} (|\bar{\theta}|M)\right]
\end{eqnarray}

\section{Explicit one loop calculation}

In this section, we will explicitly evaluate the self-energy at one
loop, using dimensional regularization. However, even before doing the
calculation, let us verify 
explicitly that the self-energy is indeed transverse at one loop, as
a check on our general result of section {\bf III}.

There are three graphs for the self-energy, at one loop, that we are
interested  in, namely, the one with the internal ghost loop, the tadpole
involving the four photon vertex and the one involving an internal  gauge
loop (see figures 1a, 1b and 1c). 
\begin{figure}[h!]
\begin{center}
\includegraphics*{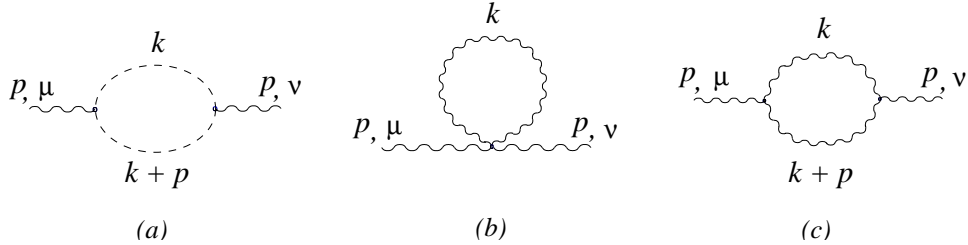}
\end{center}

\caption{One-loop diagrams which contribute to the photon-self energy. 
Wavy and dashed lines denote respectively photons and ghosts.
The external momenta on the left side is inward.} \label{fig1}
  \end{figure}
\noindent
Each of these three graphs 
has the form (see(\ref{propagator},\ref{vertex})) 
\begin{equation}
\Pi_{\mu\nu}^{(I)} = {e^{2}C^{(I)}\over 2}\, \int {d^{n}k\over
(2\pi)^{n}}\,{(1-\cos p\times k)\over k^{2} (k+p)^{2}}\,N_{\mu\nu}^{(I)};
\;\;\;\;\;\;\;\;\;\;\; I =a,b,c
\end{equation}
where
\begin{eqnarray}
C^{\rm (ghost)} & = & 1, \quad C^{\rm (tadpole)} = {1\over 4},\quad
C^{\rm (gauge)} = {1\over 2}\nonumber\\
N_{\mu\nu}^{\rm (ghost)} & = & 2 (k+p)_{\mu} k_{\nu}\nonumber\\
N_{\mu\nu}^{\rm (tadpole)} & = & - 8 (k+p)^{2}
\left[(1-n)\eta_{\mu\nu} + (1-\xi)\,\left(\eta_{\mu\nu} -
{k_{\mu}k_{\nu}\over k^{2}}\right)\right]\nonumber\\
N_{\mu\nu}^{\rm (gauge)} & = & R_{\mu\nu} + (1-\xi) S_{\mu\nu} +
(1-\xi)^{2} T_{\mu\nu}
\end{eqnarray}
where  
\begin{eqnarray}
R_{\mu\nu} & = & 2 \left[(3+n) p_{\mu}p_{\nu} + (6-4n) k_{\mu}k_{\nu}
- 2 \eta_{\mu\nu} (k^{2}+2p^{2})\right]\nonumber\\
S_{\mu\nu} & = & 2 \left[ {(k^{2}+2p\cdot k)^{2}\over
k^{2}}\,\eta_{\mu\nu} - {k^{2}+ 2p\cdot k - p^{2}\over
k^{2}}\,k_{\mu}k_{\nu} + p_{\mu}p_{\nu} - {k^{2}+ 3p\cdot k\over
k^{2}}\, (k_{\mu}p_{\nu}+k_{\nu}p_{\mu}) + (k\rightarrow k+p,
p\rightarrow -p)\right]\nonumber\\
T_{\mu\nu} & = & - 2\,{(p^{2}k_{\mu} - p\cdot k\,p_{\mu})(p^{2}
k_{\nu} - p\cdot k\,p_{\nu})\over k^{2} (k+p)^{2}}\label{gauge}
\end{eqnarray}

Thus, adding the three terms, we see that the self-energy has a
natural expansion in powers of $(1-\xi)$ ($\xi=1$ corresponds to the
Feynman gauge). The terms proportional to $(1-\xi)^{2}$ come only from
the graph with the gauge loop and its structure, as can be seen from
(\ref{gauge}), is manifestly transverse. There are two terms proportional to
$(1-\xi)$, coming from the tadpole as well as the gauge loop
diagrams. If we add them, combine the denominators using Feynman
parameters and shift the integration variable, we obtain
\begin{eqnarray}
\Pi_{\mu\nu}^{(1-\xi)} & = & {4e^{2}(1-\xi)\over (2\pi)^{n}}
\int_{0}^{1} dx (1-x) \int d^{n}k\,{(1 - \cos p\times k)\over (k^{2} +
x(1-x)p^{2})^{3}}\nonumber\\
 &  & \quad\times \left[\left((1+2x) k^{2}p^{2} - 4(1-x) (p\cdot
k)^{2} - x^{2}(3-2x) (p^{2})^{2}\right) \eta_{\mu\nu} - 2p^{2}
k_{\mu}k_{\nu}\right.\nonumber\\
 &  & \qquad\left. - \left((1+2x) k^{2} -
x^{2}(3-2x)p^{2}\right)p_{\mu}p_{\nu} + (3-2x)p\cdot k (k_{\mu}p_{\nu}
+ k_{\nu}p_{\mu})\right]
\end{eqnarray}
Contracting with $p_{\mu}$, it can be seen that this vanishes upon
symmetric integration. Therefore, the terms linear in $(1-\xi)$ are
explicitly transverse as well. In a similar manner, it can also be
checked that the terms independent of $(1-\xi)$ are
transverse. Alternatively, let us note that the terms independent of
$(1-\xi)$ would correspond to the self-energy in the Feynman gauge
($\xi=1$), which have been explicitly checked earlier to be transverse
at one loop.

Thus, we see from the structure of the graphs that they are manifestly
transverse at one loop, consistent with our all orders result. Thus,
let us parameterize the self-energy as in (\ref{selfenergy}) (with the
identification
$\bar{u}^{\mu}\equiv \bar{\theta}^{\mu} = \theta^{\mu\nu}p_{\nu}$) and write
\begin{equation}
\Pi_{\mu\nu} = A\,\left(\eta_{\mu\nu} - {p_{\mu}p_{\nu}\over
p^{2}}\right) + B\,{\bar{\theta}_{\mu}\bar{\theta}_{\nu}\over
\bar{\theta}^{2}}
\end{equation}
It follows from this that
\begin{eqnarray}
A & = & {1\over (n-2)}\,\left(\eta^{\mu\nu}\Pi_{\mu\nu} -
{\bar{\theta}^{\mu}\bar{\theta}^{\nu}\over
\bar{\theta}^{2}}\Pi_{\mu\nu}\right)\nonumber\\
B & = & {1\over (n-2)}\,\left(- \eta^{\mu\nu}\Pi_{\mu\nu} + (n-1)
{\bar{\theta}^{\mu}\bar{\theta}^{\nu}\over
\bar{\theta}^{2}}\,\Pi_{\mu\nu}\right)
\end{eqnarray}
In spite of the appearance of $(n-2)$ factors in the denominator, we
will see that these quantities are well behaved at $n=2$ (namely, two
dimensions).

Let us further expand each of these coefficients in powers of
$(1-\xi)$,
\begin{eqnarray}
A & = & A_{0} + (1-\xi) A_{1} + (1-\xi)^{2} A_{2}\nonumber\\
B & = & B_{0} + (1-\xi) B_{1} + (1-\xi)^{2} B_{2}
\end{eqnarray}
where we have used the fact that the self-energy diagrams are at most
quadratic in $(1-\xi)$. The subscripts here correspond to the power of
$(1-\xi)$ and the lowest order terms simply correspond to the
coefficients in the Feynman gauge. We can combine denominators and
integrate over the internal momenta using formulae (\ref{dimreg1}) for
the  terms
depending on $\cos p\times k$ and use the conventional formulae of dimensional
regularization for the term independent of $\theta^{\mu\nu}$. As we
have mentioned earlier, the usual results of dimensional
regularization can be obtained through the simple substitution
\begin{equation}
z^{\nu} K_{\nu} (z)\rightarrow 2^{\nu -1} \Gamma (\nu)
\end{equation}
Therefore, it is enough to use the formulae in (\ref{dimreg1}) to do
the  complete
integrals and, with a little bit of algebra, we can write the results
as 
\begin{eqnarray}
A_{i} & = & {e^{2}\pi^{n\over 2}\over (2\pi)^{n}}\,(2p^{2})^{{n\over
2}-1} \int_{0}^{1} dx\,\left(x(1-x)\right)^{{n\over 2} -2}\,
a_{i}\nonumber\\
B_{i} & = & {e^{2}\pi^{n\over 2}\over (2\pi)^{n}}\,(2p^{2})^{{n\over
2}-1} \int_{0}^{1} dx\,\left(x(1-x)\right)^{{n\over 2} -2}\,
b_{i}
\end{eqnarray}
where $i=0,1,2$ and
\begin{eqnarray}
a_{0} & = & \left(3 + 2(n-1)x -
4(n-2)x^{2}\right)\left((|\bar{\theta}|M)^{2-{n\over 2}} K_{2-{n\over
2}} (|\bar{\theta}|M) - 2^{1-{n\over 2}}\Gamma(2-{n\over
2})\right)\nonumber\\
a_{1} & = & - 2(6x^{2} - 5 x)\left((|\bar{\theta}|M)^{2-{n\over 2}}
K_{2-{n\over 2}} (|\bar{\theta}|M) - 2^{1-{n\over 2}} \Gamma(2-{n\over
2})\right)\nonumber\\
 &  & \quad + (1+4x-4x^{2})\left((|\bar{\theta}|M)^{3-{n\over 2}}
K_{3-{n\over 2}} (|\bar{\theta}|M) - 2^{2-{n\over 2}} \Gamma(3-{n\over
2})\right)\nonumber\\
a_{2} & = & {1\over 4} \left((|\bar{\theta}|M)^{3-{n\over 2}}
K_{3-{n\over 2}} (|\bar{\theta}|M) - 2^{2-{n\over 2}} \Gamma(3-{n\over
2})\right)\nonumber\\
b_{0} & = & - 4 (n-2)^{2} x(1-x) \left((|\bar{\theta}|M)^{1-{n\over
2}} K_{1-{n\over 2}} (|\bar{\theta}|M) - 2^{-{n\over 2}}
\Gamma(1-{n\over 2})\right)\nonumber\\
 &  &  \quad +\left((6-4n) x(1-x) +
1-2x^{2}\right)\left((|\bar{\theta}|M)^{2-{n\over 2}} K_{2-{n\over 2}}
(|\bar{\theta}|M) - 2^{1-{n\over 2}} \Gamma(2-{n\over
2})\right)\nonumber\\
b_{1} & = & 2x\left[ (n-4) \left((|\bar{\theta}|M)^{2-{n\over 2}}
K_{2-{n\over 2}} (|\bar{\theta}|M) - 2^{1-{n\over 2}} \Gamma(2-{n\over
2})\right) + \left((|\bar{\theta}|M)^{3-{n\over 2}} K_{3-{n\over 2}}
(|\bar{\theta}|M) - 2^{2-{n\over 2}} \Gamma(3-{n\over
2})\right)\right]\nonumber\\
b_{2} & = & -{1\over 4}\left[ (n-6)\left((|\bar{\theta}|M)^{3-{n\over
2}} K_{3-{n\over 2}} (|\bar{\theta}|M) - 2^{2-{n\over 2}}
\Gamma(3-{n\over 2})\right) + \left((|\bar{\theta}|M)^{4-{n\over 2}}
K_{4-{n\over 2}} (|\bar{\theta}|M) - 2^{3-{n\over 2}} \Gamma(4-{n\over
2})\right)\right]
\end{eqnarray}

Combining all the factors, we obtain
(after rotation to the Minkowski space)
\begin{eqnarray}\label{B}
A & = & {e^{2}\pi^{n\over 2}\over (2\pi)^{n}}\,(2p^{2})^{{n\over 2}-1}
\int_{0}^{1} dx (x(1-x))^{{n\over 2}-2}
\left[\left(3+2(n-1)x-4(n-2)x^{2}-2(1-\xi)x(6x-5)\right)\right.\nonumber\\
 &  & \times\left((|\bar{\theta}|M)^{2-{n\over
2}} K_{2-{n\over 2}} (|\bar{\theta}|M) - 2^{1-{n\over 2}}
\Gamma(2-{n\over 2})\right)\nonumber\\
 &  & \qquad \left. -(1-\xi)\left((1+4x-4x^{2}) - {(1-\xi)\over
4}\right)\left((|\bar{\theta}|M)^{3-{n\over 2}} K_{3-{n\over 2}}
(|\bar{\theta}|M) - 2^{2-{n\over 2}} \Gamma(3-{n\over
2})\right)\right]\nonumber\\
B & = & {e^{2}\pi^{n\over 2}\over (2\pi)^{n}}\,(2p^{2})^{{n\over 2}-1}
\int_{0}^{1} dx (x(1-x))^{{n\over 2}-2} \left[-4(n-2)^{2}
x(1-x)\left((|\bar{\theta}|M)^{1-{n\over 2}} K_{1-{n\over 2}}
(|\bar{\theta}|M) - 2^{-{n\over 2}} \Gamma(1-{n\over
2})\right)\right.\nonumber\\
 &  & \qquad +\left((6-4n) x(1-x) + 1-2x^{2}
+2(1-\xi)(n-4)x\right)\left((|\bar{\theta}|M)^{2-{n\over 2}}
K_{2-{n\over 2}} (|\bar{\theta}|M) - 2^{1-{n\over 2}} \Gamma(2-{n\over
2})\right)\nonumber\\
 &  & \qquad + (1-\xi)\left(2x - (n-6){(1-\xi)\over
4}\right)\left((|\bar{\theta}|M)^{3-{n\over 2}} K_{3-{n\over 2}}
(|\bar{\theta}|M) - 2^{2-{n\over 2}} \Gamma(3-{n\over
2})\right)\nonumber\\
 &  & \qquad\left. -
{(1-\xi)^{2}\over 4}\left((|\bar{\theta}|M)^{4-{n\over 2}}
K_{4-{n\over 2}} (|\bar{\theta}|M) - 2^{3-{n\over 2}} \Gamma(4-{n\over
2})\right)\right],
\end{eqnarray}
where $M^2 = -x(1-x) p^2$.
We note that the integration over the Feynman parameter, $x$, can
be done in closed form, for both the coefficients, in terms of
generalized hypergeometric functions. However, the result is not very
illuminating and, therefore, we do not give the details here.

This, therefore, determines the photon self-energy, at one loop, in a
general covariant gauge in any dimension. There are several things to
note from this result. First of all, the coefficients are, in general,
dependent on the gauge fixing parameter, $\xi$, as is the case in
commutative QCD at zero temperature (in commutative QED, these
coefficients are gauge independent). Second, in spite of the
complicated  structure in $B$, all  the
$\Gamma$ terms cancel out exactly, when the integration over the Feynman
parameter is carried out. This is true in any dimension and for any
value of $\xi$ and this is an important result. For, it says that
there is no ultraviolet divergence in the coefficient of $B$ in any dimension 
in a general covariant gauge. Therefore, all the ultraviolet
divergences are contained in
$A$ and can be subtracted by the usual wave function renormalization
counterterms. We do not need any counterterm with a new structure in
the non-commutative QED, which would have rendered the theory
unrenormalizable. 

Second, when $n=2$, $\theta^{\mu\nu}=0$ (since there is only one
space direction). In two dimensions, the theory is
ultraviolet  finite by power counting and gauge invariance, therefore,
the $\theta^{\mu\nu}\rightarrow 0$ limit can be taken smoothly in our
results.  In
this limit, of course, $(1-\cos p\times k)\rightarrow 0$ so that we will
expect these structures to vanish when $n=2$. This can be explicitly
checked in the following way. Note that when the integral is
convergent, as noted in (\ref{convgt}),
\[
\lim_{z\rightarrow 0}\,z^{\nu} K_{\nu} (z) = 2^{\nu -1} \Gamma(\nu)
\]
In such a case, every term inside the parenthesis will cancel pairwise
to give a vanishing result.

When $n=4-2\epsilon$, the terms in $A$, related to the $\Gamma$
functions, have the explicit form
\begin{equation}
A_{\rm planar} = - {2e^{2}p^{2}\over (4\pi)^{2}}\,\left[\left({13\over
6} - {\xi\over 2}\right) \left({1\over \epsilon} - \log {-p^{2}\over
4\pi \mu^{2}} - \gamma\right) + {31\over 9} - (1-\xi) +
{(1-\xi)^{2}\over 4}\right]
\end{equation}
whereas the leading order term, coming from the Bessel functions, as
$|\bar{\theta}|\rightarrow 0$, takes the
simple form
\begin{equation}
A_{\rm non-planar} = - {2e^{2}p^{2}\over
(4\pi)^{2}}\,\left[\left({13\over 6} - {\xi\over 2}\right) 
\log(-p^2 |\bar{\theta}|^{2}) + O(|\bar{\theta}|^{2})\right]
\end{equation}
Let us note that the $\log (-p^{2})$ terms precisely cancel between the
planar and the non-planar terms. This is a general feature that is
completely parallel with the studies at finite temperature (see next
section for more details). As for the
coefficient $B$, we have already seen that all the $\Gamma$ terms
cancel out in any dimension, so that
\begin{equation}
B_{\rm planar} = 0
\end{equation}
The leading order term coming from the Bessel functions, as
$|\bar{\theta}|\rightarrow 0$, in four dimensions has the form
\begin{equation}
B_{\rm non-planar} = - {e^{2}\over
16\pi^{2}}\,\left[\left({32\over |\bar{\theta}|^{2}} - {4p^{2}\over
3}\right) + O(|\bar{\theta}|^{2})\right]\label{sing}
\end{equation}
which agrees with the results in references  \cite{Hayakawa:1999zf,Ruiz:2000hu}
for QED in 4-dimensions. (We remark here that the leading
order contribution, as $|\bar\theta|\rightarrow 0$, comes from the 
$(|\bar{\theta}|M)^{1-{n\over 2}}
K_{1-{n\over 2}} (|\bar{\theta}|M)$ term in (\ref{B}), whose
coefficient is  gauge independent.) 

\section{Imaginary part of the self-energy}

As is well known, non-commutative theories do not have a unique
$\theta^{\mu\nu}\rightarrow 0$ limit. This non-analyticity is quite
analogous to the behavior in thermal field theories, where the
amplitudes become non-analytic at the origin in the energy-momentum
plane. In the case of thermal field theories, there is a physical
reason for such a non-analyticity, namely, at finite temperature,
new channels of reaction  develop leading to new thermal
branch cuts, and this leads to the
non-analyticity. Correspondingly, it would be interesting to ask if
the non-commutative QED theory develops any new $\theta^{\mu\nu}$
dependent imaginary parts in the amplitude.

To this end, let us start by looking at the self-energy in a
non-commutative $\phi^{3}$ theory in $n$ dimensions. We choose the
scalar field to be massless so as to keep the parallel with
non-commutative QED. In this case, the basic integral for the
non-planar part of the self-energy has the form (in Euclidean space)
\begin{equation}
I_{\rm non-planar} \sim {1\over (2\pi)^{n}} \int_{0}^{1} dx \int
d^{n}k\, {e^{i\bar{\theta}\cdot k}\over (k^{2} + M^{2})^{2}}
\end{equation}
where we are neglecting some overall
multiplicative factors for simplicity. This integral can be evaluated
using (\ref{dimreg1}) and gives  (after rotation to Minkowski space)
\begin{equation}
I_{\rm non-planar} \sim {\pi^{n\over 2}\over (2\pi)^{n}}\,\int_{0}^{1}
dx\,{2^{{n\over
2} -1}\over (M^{2})^{2-{n\over 2}}}\,(|\bar{\theta}|M)^{2-{n\over 2}}
K_{2-{n\over 2}} (|\bar{\theta}|M)\label{scalar}
\end{equation}
where, in the Minkowski space $M^{2} = - x(1-x) p^{2}$. The planar
part of the self-energy,  on the other hand, has no
exponential factor in the integrand and the result is 
\begin{equation}
I_{\rm planar} \sim {\pi^{n\over 2}\over (2\pi)^{n}}\,\int_{0}^{1}
dx\, {\Gamma(2-{n\over 2})\over (M^{2})^{2-{n\over 2}}}\label{scalar1}
\end{equation}

We note that for $n<4$, the integral is ultraviolet
convergent and, in the limit $|\bar\theta|\rightarrow 0$, (\ref{scalar})
yields the result in (\ref{scalar1}). In such a
case, we do not expect any non-analyticity in the theory. On the
other hand, it is for $n>4$ that ultraviolet divergences are
present leading to IR/UV mixing, which is the main reason for the
non-analytic behavior as $\theta^{\mu\nu}\rightarrow 0$. So, let us
analyze the imaginary parts of these amplitudes for $n>4$. First, let
us note that if we set $n=4+2\ell-2\epsilon$ with $\ell$ an integer
and $\epsilon$ infinitesimal,
then,  with some algebra, the planar term in (\ref{scalar1}) becomes
\begin{equation}
I_{\rm planar} \sim {\pi^{n\over 2}\over (2\pi)^{n}}\,\int_{0}^{1}
dx\,M^{2\ell}\,{(-1)^{\ell}\over \ell !}\left({1\over \epsilon} + \log
{\mu^{2}\over M^{2}} + \cdots\right)
\end{equation}
Here $\mu$ is the scale of dimensional regularization and we see that,
since $M^{2}=-x(1-x)p^{2}$, for $p^{2}>0$, the logarithm will lead to
an imaginary part.

In the evaluation of the non-planar term, on the other hand, we can
safely set $\epsilon=0$ (it has no poles) and the Bessel function can
be expanded for small $|\bar{\theta}|$ to give
\begin{eqnarray}
I_{\rm non-planar} &\sim & {\pi^{n\over 2} \over  (2\pi)^{n}}\,\int_{0}^{1}
dx\left[(|\bar{\theta}|)^{-2\ell}\,\sum_{k=0}^{\ell-1}
(|\bar{\theta}|M)^{2k} {(-1)^{k}(\ell-k -1)!
2^{2(\ell-k)}\over k!}\right.\nonumber\\
 &  & \left. + (-1)^{\ell+1} M^{2\ell}
\sum_{k=0}^{\infty} {(|\bar{\theta}|M)^{2k}\over 2^{2k} k!
(\ell+k)!}\left(\log {|\bar{\theta}|^{2}M^{2}\over 4} - \psi(k+1) -
 \psi(\ell+k+1)\right)\right]\label{expansion}
\end{eqnarray}
where $\psi(x)$ is the Euler psi function  \cite{gradshteyn}.
Let us note that, unlike
the real part, the imaginary part of (\ref{expansion}), which comes
from the $\log M^{2}$ terms, is a well behaved function in the limit
$|\bar{\theta}|\rightarrow 0$.  
The leading imaginary part which arises from this when
$|\bar{\theta}|\rightarrow 0$, comes from the term
\begin{equation}
I_{\rm non-planar}^{\rm (leading\ log)} \sim {\pi^{n\over 2}\over
(2\pi)^{n}}\, \int_{0}^{1} dx M^{2\ell} {(-1)^{\ell}\over  
\ell !}\,\log {4\over |\bar{\theta}|^{2}M^{2}}
\end{equation}
It is interesting to note that the $\log M^{2}$ term in the planar and
the non-planar terms have the same coefficient. This is very much like
the behavior of $\log T$ terms in thermal field theories (Here, $T$ is
the temperature). Namely,
while there is no direct relation between the ultraviolet divergence
in a field theory and powers of $T$, the coefficient of $\log T$
coincides with that of the pole ${1\over \epsilon}$ \cite{Brandt:1999gm}. 
Here, too, the same behavior seems to arise.

The same discussion carries over to non-commutative QED, where the
imaginary parts of the non-planar terms in (\ref{B}) may be evaluated
using the relation (for $M^2 < 0 $)
\begin{equation}
{\rm Im}\, K_{\ell} (|\bar{\theta}|M) = (-1)^{\ell+1}\,{\pi\over 2}\,
J_{\ell} (|\bar{\theta}||M|)\label{bessel}
\end{equation}
Let us
note that, in the case of QED, the planar and the non-planar terms
come with opposite sign because of the factor $(1-\cos p\times k)$. As
a result, to leading order, the imaginary parts cancel in the
self-energy. However, there are higher order terms in the expansion of
the Bessel function in (\ref{bessel}), which can contribute an
imaginary  part to
the self-energy. When $|\bar{\theta}|\rightarrow 0$, however, these vanish
quite rapidly. On the other hand, for finite $|\bar{\theta}|$, these
imaginary  parts
are present and are, in fact, necessary for unitarity to hold
 \cite{Gomis:2000zz,Alvarez-Gaume:2001ka,Seiberg:2000ms,Bassetto:2001vf}. 
As we have
seen in the last section, the coefficients $A$ and $B$ in the
self-energy are gauge dependent. Therefore, we conclude that the
imaginary parts coming from the higher order terms in the Bessel
function will also become gauge dependent. This is slightly surprising
in that we would expect the imaginary part of an amplitude to be
related to a physical cross section, which has to be gauge
independent. The puzzle is resolved by noting that the physical
process to which the photon self-energy can
contribute is the electron-electron scattering amplitude (see figure 2).
\begin{figure}[h!]
\begin{center}
\includegraphics*{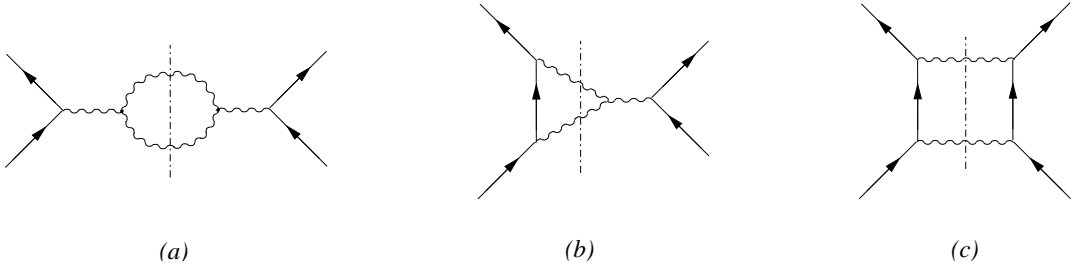}
\end{center}

\caption{Examples of one-loop diagrams which contribute to the
  electron-electron scattering in non-commutative QED.}\label{fig2}
  \end{figure}
Although the imaginary part of the self-energy graph is  
gauge dependent in this theory, it turns out that the other diagrams 
are also gauge dependent so that the sum of all such contributions can 
become gauge independent.

\section{Gauge independence of the poles of the propagator}

As we have already seen in section {\bf V}, the coefficient functions
$A$ and $B$ in the photon self-energy are gauge dependent. Therefore,
it is natural to ask what happens to the poles of the photon
propagator in such a theory. In what follows, we will show that, in
spite of the gauge dependence of the self-energy, the poles in the
photon propagator are gauge independent to all orders.

Following our discussions in section {\bf III}, we note that the
general structure of the complete two point function for the photon
(to all orders) has the form (with $\bar{u}^{\mu}\equiv
\bar{\theta}^{\mu}$) 
\begin{equation}
\Gamma^{\mu\nu} = (p^{2} + A) \left(\eta^{\mu\nu} -
{p^{\mu}p^{\nu}\over p^{2}}\right) +
B\,{\bar{\theta}^{\mu}\bar{\theta}^{\nu}\over \bar{\theta}^{2}} +
{p^{\mu}p^{\nu}\over \xi}
\end{equation}
Drawing  from previous experience at finite temperature \cite{kobes:1990xf}, 
we note that, for the
purpose  of analyzing the gauge independence of the poles of the
propagator,  it is better to rewrite the two point function as 
\begin{equation}
\Gamma^{\mu\nu} = (p^{2} + A) \left(\eta^{\mu\nu} -
{p^{\mu}p^{\nu}\over p^{2}} -
{\bar{\theta}^{\mu}\bar{\theta}^{\nu}\over \bar{\theta}^{2}}\right) +
(p^{2} + A + B)\,{\bar{\theta}^{\mu}\bar{\theta}^{\nu}\over
\bar{\theta}^{2}} + {p^{\mu}p^{\nu}\over \xi}\label{decomp1}
\end{equation}
It is easy to determine from this that the exact propagator for the
photon has the form
\begin{equation}
- D_{\mu\nu} = \left(\eta_{\mu\nu} - {p_{\mu}p_{\nu}\over p^{2}} -
{\bar{\theta}_{\mu}\bar{\theta}_{\nu}\over
\bar{\theta}^{2}}\right)\,{1\over p^{2}+A} +
{\bar{\theta}_{\mu}\bar{\theta}_{\nu}\over \bar{\theta}^{2}}\,{1\over
p^{2}+A+B} + \xi\,{p_{\mu}p_{\nu}\over (p^{2})^{2}}
\end{equation}
This  reduces to the tree level
propagator when $A=B=0$ and  the exact propagator satisfies the 't
Hooft identity, $p^{\mu}p^{\nu}D_{\mu\nu}=-\xi$.

The main reason for rewriting the two point function in the form
(\ref{decomp1}) is that
the two structures
\begin{eqnarray}
P_{\mu\nu} & = & \left(\eta_{\mu\nu} - {p_{\mu}p_{\nu}\over p^{2}} -
{\bar{\theta}_{\mu}\bar{\theta}_{\nu}\over
\bar{\theta}^{2}}\right)\nonumber\\
Q_{\mu\nu} & = & {\bar{\theta}_{\mu}\bar{\theta}_{\nu}\over
\bar{\theta}^{2}}
\end{eqnarray} 
are orthogonal, transverse projection operators satisfying
\begin{eqnarray}
P_{\mu\nu} P^{\nu}_{\lambda} & = & P_{\mu\lambda}\nonumber\\
Q_{\mu\nu}Q^{\nu}_{\lambda} & = & Q_{\mu\lambda}\nonumber\\
P_{\mu\nu}Q^{\nu}_{\lambda} & = & 0 = p^{\mu}P_{\mu\nu} =
p^{\mu}Q_{\mu\nu} 
\end{eqnarray}
which will be quite useful in the following analysis. Furthermore, in
this way of writing, we see clearly that the propagator has two
independent poles at $p^{2}+A=0,\, p^{2}+A+B=0$. (The pole in the
longitudinal part has a gauge dependent residue and
is clearly unphysical. Note that, at finite  temperature,
the physical poles are related to the Debye and the plasmon masses.)

The gauge dependence of the two point function and, therefore, of the
poles of the propagator can be analyzed through Nielsen identities,
which we will derive in the appendix. For the present, let us simply
note that the change in the two point function, under a change in the
gauge fixing parameter, can be written as (in momentum space)
\begin{equation}
{\partial\Gamma^{\mu\nu}\over \partial\xi} =
\left[\Gamma^{\mu\lambda}X_{\lambda}^{\nu} + \Gamma^{\nu\lambda}
X_{\lambda}^{\mu}\right]\label{var}
\end{equation}
where the quantity $ X_{\lambda}^{\mu}$ is described in the appendix.
Taking the projection of (\ref{var}) with
$P_{\nu\mu}$, we obtain,
\begin{equation}
{\partial (p^{2} + A)\over \partial \xi} = {2 (p^{2} + A)\over
(n-2)}\,P_{\mu\nu} X^{\nu\mu}
\end{equation}
Similarly, taking the projection of (\ref{var}) with $Q_{\mu\nu}$, we obtain,
\begin{equation}
{\partial (p^{2} + A + B)\over \partial \xi} = 2 (p^{2} + A +
B)\,Q_{\mu\nu} X^{\nu\mu}
\end{equation}

These two equations are quite interesting as they say that, since
$(p^{2} + A)$ as well as $(p^{2} + A + B)$ change homogeneously as we
change the gauge fixing parameter, the zeroes of these functions are
gauge independent. Correspondingly, the poles of the propagator are
gauge independent. Namely, even
though the photon two point function is gauge dependent, to all orders, the
poles of the photon propagator are gauge independent (in any
dimension). Let us note here that an important consequence of this
property  is that the most infrared singular term in the above equations
must be gauge independent. Otherwise, the poles of the
propagator would not have a gauge independent location. In fact, by
explicit calculation, we find that this term appears in $B$ in the
gauge independent form
\begin{equation}
B^{\rm singular} = - {e^{2} (n-2)^{2}\over 2 \pi^{n\over
2}}\,\left({1\over |\bar{\theta}|}\right)^{n-2}
\end{equation}
which clearly vanishes for $n=2$ and which can be compared with
(\ref{sing}) for $n=4$.

\section{Conclusion}

In this paper, we have studied the contributions of gauge and ghost
loops to the photon self-energy in non-commutative QED, in a general
covariant gauge {\it and} in any dimension (The fermion contributions have been
studied earlier). We have shown that, to all orders, the self-energy
is transverse and we have explicitly evaluated the one-loop graphs,
which verify this. Our calculations have used dimensional
regularization, which we have generalized to non-commutative
theories. The explicit calculation shows that there are no new kinds of
ultraviolet divergences coming from these diagrams so that the theory
is renormalizable \cite{Bichl:2001cq}. Furthermore, the
imaginary parts coming from these graphs cancel identically in the
infrared limit, although away from the infrared limit, the self-energy
does have $\theta$-dependent imaginary parts which are necessary for
unitarity. Since the photon self-energy is gauge dependent, we use the
Nielsen identity to show that the poles of the photon propagator are
gauge independent to all orders. Generally, the $\theta$-dependent
infrared divergent terms that arise, to one-loop order, 
in non-commutative theories have
inappropriate behavior. However, since we do not find any imaginary
part associated with such a non-analyticity, it suggests that such
behavior may not, in fact, be physical. 
Drawing from studies of non-commutative scalar
models \cite{Fischler:2000fv,Gubser:2000cd,Griguolo:2001wg}, which make
use of techniques developed in connection with
thermal field theories, we conjecture that a resummation to all orders
may eliminate the unphysical infrared singularities in non-commutative QED.

\vskip .5cm
\begin{acknowledgments}  
We would  like to thank Professor J. C. Taylor for many helpful discussions. 
This work was supported in part by US DOE grant number
DE-FG-02-91ER40685 and by CNPq and FAPESP, Brazil.
\end{acknowledgments}  

\appendix*

\section{The Nielsen identity}

In this appendix, we will derive the Nielsen identity used in section
{\bf VII} to prove the gauge independence of the poles of the
propagator. 

Let us start with the action for non-commutative QED given in
(\ref{action}), where we
write  the gauge fixing term as
\begin{equation}
S_{\rm gf} = \int d^{n}x\,\left({\xi\over 2} F\star F - F\star
(\partial\cdot A)\right)
\end{equation}
Here $F$ is an auxiliary field whose equation of motion gives
\begin{equation}
F = {1\over \xi} (\partial\cdot A)
\end{equation}
which we will use later in the analysis.
However, it is more convenient to begin with the auxiliary field formulation of
the gauge fixing, since it allows the BRST transformations of the
theory to close off-shell. The BRST transformations for
non-commutative QED, in this formulation, become
\begin{eqnarray}
\delta A_{\mu} & = & \omega D_{\mu}c = \omega \left(\partial_{\mu}c -
ie \left[A_{\mu},c\right]_{\rm MB}\right)\nonumber\\
\delta c & = & - {\omega\over 2}\,c\star c\nonumber\\
\delta \bar{c} & = & - \omega F\nonumber\\
\delta F & = & 0
\end{eqnarray}
Here $\omega$ is an anti-commuting space-time independent parameter
and the action $S$, which includes gauge fixing and ghosts, is
invariant under these transformations.

Let us now add to our action source terms
\begin{equation}
S_{\rm source} = \int d^{n}x\,\left(J^{\mu}\star A_{\mu} + J\star F +
i (\bar{\eta}\star c - \bar{c}\star \eta) + K^{\mu}\star D_{\mu}c +
L\star (-{1\over 2}c\star c) + H\star ({1\over 2} \bar{c}\star F)\right)
\end{equation}
Here, we have the usual sources for the fields, sources for the
composite BRST variations and finally, we have added one extra source
(the last term) whose meaning will become clear shortly. The action
involving the sources is not invariant under the BRST transformations
and gives
\begin{equation}
\delta S_{\rm source} = \omega \int d^{n}x\,\left(J^{\mu}\star
D_{\mu}c + i \bar{\eta} ({1\over 2} c\star c) + i F\star \eta + H\star
({1\over 2} F\star F)\right)\label{source}
\end{equation}

Let us next define the generating functional as
\begin{equation}
Z = e^{iW} = \int [{\cal D}\varphi]\, e^{i(S + S_{\rm source})}
\end{equation}
where $\varphi$ represents all the fields being integrated and the
generating functional depends only on the sources. It is clear now
that, under a BRST field redefinition inside the path integral, the
generating functional will not change, since the sources are
unaffected by such a transformation. This leads to
\begin{eqnarray}
\delta Z & = & 0\nonumber\\
\int [{\cal D}\varphi]\,(\delta S + \delta S_{\rm source})\,e^{i(S +
S_{\rm source})} & = & 0
\end{eqnarray}
Since $S$ is invariant, using the form of $\delta S_{\rm source}$ from
(\ref{source}), we obtain,
\begin{equation}
\int [{\cal D}\varphi]\,\left(\int d^{n}z\,H(z)\star {\partial {\cal L}\over
\partial \xi}\right) e^{i(S+S_{\rm source})} = - \int
d^{n}z\,\left(J^{\mu}\star {\delta W\over \delta K^{\mu}(z)} -
i\bar{\eta}(z)\star {\delta W\over \delta L(z)} + i {\delta W\over
\delta J(z)}\star \eta(z)\right)
\end{equation}
Taking the derivative of this with respect to $H(x)$, setting it to zero
and integrating over $x$, we obtain
\begin{equation}
{\partial W\over \partial \xi} = - \int
d^{n}x\,d^{n}z\,\left.\left(J^{\mu}(z)\star {\delta^{2}W\over \delta
H(x)\delta K^{\mu}(z)} + i\bar{\eta}(z)\star {\delta^{2} W\over \delta
H(x) \delta L(z)} + i {\delta^{2} W\over \delta H(x) \delta J(z)}\star
\eta(z)\right)\right|_{H=0}\label{iden}
\end{equation}

We can now define the effective action, $\Gamma$, through the Legendre
transformation
\begin{equation}
\Gamma = W - \int d^{n}x\,\left(J^{\mu}\star A_{\mu} + J\star F + i
(\bar{\eta}\star c - \bar{c}\star \eta)\right)
\end{equation}
Then, the identity in (\ref{iden}) can be written as
\begin{equation}
{\partial \Gamma\over \partial \xi} =  \int d^{n}x\,d^{n}z\,\left.\left(
{\delta \Gamma\over \delta A_{\lambda}(z)}\star {\delta^{2}\Gamma\over
\delta H(x)\delta K^{\lambda}(z)} - {\delta\Gamma\over \delta
{c}(z)}\star {\delta^{2}\Gamma\over \delta H(x)\delta L(z)} +
{\delta F(z)\over \delta H(x)}\star {\delta\Gamma\over \delta
\bar{c}(z)}\right)\right|_{H=0}
\end{equation}
Taking the second derivative with respect to $A_{\mu}(x)$ and
$A_{\nu}(y)$, setting $F={1\over \xi}(\partial\cdot A)$ and setting
all other fields equal to zero, we obtain
\begin{equation}
{\partial\over \partial \xi} {\delta^{2}\Gamma\over \delta
A_{\mu}(x)\delta A_{\nu}(y)} = \int
d^{n}\omega\,d^{n}z\left.\left({\delta^{2}\Gamma\over \delta A_{\mu}(x)\delta
A_{\lambda}(z)}\star {\delta^{3}\Gamma\over \delta H(\omega)\delta
A_{\nu}(y)\delta K^{\lambda}(z)} + {\delta^{2}\Gamma\over \delta
A_{\nu}(y)\delta 
A_{\lambda}(z)}\star {\delta^{3}\Gamma\over \delta H(\omega)\delta
A_{\mu}(x)\delta K^{\lambda}(z)}\right)\right|
\end{equation}
Here, the restriction is understood as setting  $F = {1\over \xi}
(\partial\cdot A)$ and, then, setting all the fields equal to zero,
after taking the functional derivatives. This is the identity
used in  section {\bf VII} (see (\ref{var})), where we have
identified,
\begin{equation}
X_{\lambda}^{\mu} = 
\left.{\delta^{3}\Gamma\over \delta H \delta A_{\mu} \delta
K^{\lambda}}\right|
\end{equation}
A graphical representation for $X_{\lambda}^{\mu}$ to lowest orders is
shown in figure 3.
\begin{figure}[h!]
\begin{center}
\includegraphics*{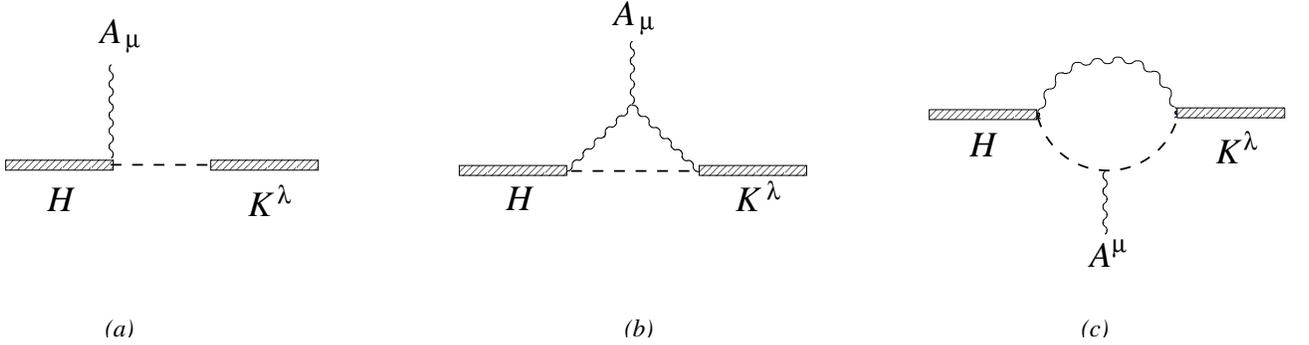}
\end{center}

\caption{The diagrammatic expression of Eq. (A13): The lowest order
term (a) and the one-loop contributions (b and c).}\label{fig3}
  \end{figure}
\appendix

\end{document}